\documentclass{wscpaperproc}
\usepackage{latexsym}
\usepackage{graphicx}
\usepackage{mathptmx}
\usepackage[T1]{fontenc}

\usepackage{amsmath}
\usepackage{amsfonts}
\usepackage{amssymb}
\usepackage{amsbsy}
\usepackage{amsthm}

\usepackage{listings}
\usepackage{xcolor}

\newcommand{\placeholder}[1]{\textbf{\texttt{\textless #1\textgreater}}}

\usepackage[pdftex,colorlinks=true,urlcolor=blue,citecolor=black,anchorcolor=black,linkcolor=black]{hyperref}

\newtheoremstyle{wsc}
{3pt}
{3pt}
{}
{}
{\bf}
{}
{.5em}
{}

\theoremstyle{wsc}

\begin{document}

%
%

\pagestyle{fancyplain}

\thispagestyle{plain}
\firstPageHead{}

\chead{\fancyplain{}{\itshape Giabbanelli}}

\rhead{}
\cfoot{}
\renewcommand{\headrulewidth}{0pt} 

\makeatletter
\let\@internalcite\cite
\def\cite{\def\@citeseppen{-1000}%
    \def\@cite##1##2{(##1\if@tempswa , ##2\fi)}%
    \def\citeauthoryear##1##2##3{##1 ##3}\@internalcite}
\def\citeNP{\def\@citeseppen{-1000}%
    \def\@cite##1##2{##1\if@tempswa , ##2\fi}%
    \def\citeauthoryear##1##2##3{##1 ##3}\@internalcite}
\def\citeN{\def\@citeseppen{-1000}%
    \def\@cite##1##2{##1\if@tempswa, ##2)\else{}\fi}%
    \def\citeauthoryear##1##2##3{##1 (##3)}\@citedata}
\def\citeA{\def\@citeseppen{-1000}%
    \def\@cite##1##2{(##1\if@tempswa , ##2\fi)}%
    \def\citeauthoryear##1##2##3{##1}\@internalcite}
\def\citeANP{\def\@citeseppen{-1000}%
    \def\@cite##1##2{##1\if@tempswa , ##2\fi}%
    \def\citeauthoryear##1##2##3{##1}\@internalcite}
\def\shortcite{\def\@citeseppen{-1000}%
    \def\@cite##1##2{(##1\if@tempswa , ##2\fi)}%
    \def\citeauthoryear##1##2##3{##2 ##3}\@internalcite}
\def\shortciteNP{\def\@citeseppen{-1000}%
    \def\@cite##1##2{##1\if@tempswa , ##2\fi}%
    \def\citeauthoryear##1##2##3{##2 ##3}\@internalcite}
\def\shortciteN{\def\@citeseppen{-1000}%
    \def\@cite##1##2{##1\if@tempswa, ##2\else{}\fi}%
    \def\citeauthoryear##1##2##3{##2 (##3)}\@citedata}
\def\shortciteA{\def\@citeseppen{-1000}%
    \def\@cite##1##2{(##1\if@tempswa , ##2\fi)}%
    \def\citeauthoryear##1##2##3{##2}\@internalcite}
\def\shortciteANP{\def\@citeseppen{-1000}%
    \def\@cite##1##2{##1\if@tempswa , ##2\fi}%
    \def\citeauthoryear##1##2##3{##2}\@internalcite}
\def\citeyear{\def\@citeseppen{-1000}%
    \def\@cite##1##2{(##1\if@tempswa , ##2\fi)}%
    \def\citeauthoryear##1##2##3{##3}\@citedata}
\def\citeyearNP{\def\@citeseppen{-1000}%
    \def\@cite##1##2{##1\if@tempswa , ##2\fi}%
    \def\citeauthoryear##1##2##3{##3}\@citedata}
%
%
%
\def\@citedata{%
    \@ifnextchar [{\@tempswatrue\@citedatax}%
                  {\@tempswafalse\@citedatax[]}%
}

\def\@citedatax[#1]#2{%
\if@filesw\immediate\write\@auxout{\string\citation{#2}}\fi%
  \def\@citea{}\@cite{\@for\@citeb:=#2\do%
    {\@citea\def\@citea{, }\@ifundefined
       {b@\@citeb}{{\bf ?}%
       \@warning{Citation `\@citeb' on page \thepage \space undefined}}%
{\csname b@\@citeb\endcsname}}}{#1}}%

%
\def\@citex[#1]#2{%
\if@filesw\immediate\write\@auxout{\string\citation{#2}}\fi%
  \def\@citea{}\@cite{\@for\@citeb:=#2\do%
    {\@citea\def\@citea{; }\@ifundefined
       {b@\@citeb}{{\bf ?}%
       \@warning{Citation `\@citeb' on page \thepage \space undefined}}%
{\csname b@\@citeb\endcsname}}}{#1}}%

%
\def\@biblabel#1{}
\makeatother



\newdimen\bibindent
\bibindent=0.0em
\def\thebibliography#1{\section*{\refname}\list
   {}{\settowidth\labelwidth{[#1]}
   \leftmargin\parindent
   \itemindent -\parindent
   \listparindent \itemindent
   \itemsep 0pt
   \parsep 0pt}
   \def\newblock{}
   \sloppy
   \sfcode`\.=1000\relax}


\setlength{\baselineskip}{12.7pt}

\title{Text-to-Image Generative AI for Modeling and Simulation:\\Methods, Opportunities, and Applications}

\author{\begin{center}Philippe J. Giabbanelli\textsuperscript{1}\\
[11pt]
\textsuperscript{1}National Center for Collaboration in Medical Modeling \& Simulation and Office of Enterprise Research \& Innovation, Old Dominion University, Suffolk, VA, USA\\
\end{center}
}

\maketitle

\vspace{-12pt}

\section*{ABSTRACT}
Text-to-image generation is a form of generative artificial intelligence (GenAI) that converts textual descriptions into images. Most applications of GenAI in modeling and simulation (M\&S) have focused on large language models for documentation, coding, or explanation. By contrast, the potential of image generation remains largely unexplored. This tutorial introduces text-to-image generation to the M\&S community and details how it can support several M\&S tasks, including communicating conceptual models, visualizing simulation outcomes, generating educational materials, and interfacing heterogeneous models in multi-scale simulations. The tutorial combines conceptual guidance with practical workflows, explaining how modern image generators operate, how prompts and simulation outputs can be translated into visual scenes, and how practitioners can integrate these tools into reproducible local pipelines. By focusing on transferable principles rather than specific tools, the tutorial equips M\&S practitioners with the knowledge needed to evaluate, adopt, and adapt text-to-image generation in their simulation workflows.

\section{INTRODUCTION}
\label{sec:intro}

Generative Artificial Intelligence (GenAI) has become a familiar sight for modelers. Large Language Models (LLMs) have been used across the Modeling \& Simulation (M\&S) pipeline~\cite{giabbanelli2023gpt}. Applications naturally include stages that are text-heavy, such as transforming text into a model specification (e.g., Agent-Based Models in~\shortciteNP{onggo2025ai}, System Dynamics in~\citeNP{schoenberg2026building}) or explaining simulation outputs to stakeholders via executive summaries~\cite{gandee2025faithful} or narrated journeys of simulated agents~\shortcite{giabbanelli2025promoting}. There is also visible interest in using the coding abilities of LLMs to facilitate implementations as part of the M\&S process~\shortcite{jackson2024natural} or to facilitate the independent reuse of models by others~\cite{monks2025unlocking}. LLMs have even been used as `experts' to guide model construction~\cite{schuerkamp2025guiding} or validation~\shortcite{moltner2025creation}. However, \emph{GenAI is not limited to LLMs}. As shown in a comprehensive review at the Winter Simulation Conference (WSC) by~\citeN{feldkamp2025use}, GenAI also covers several  techniques such as Generative Adversarial Networks (GANs) and Variational Autoencoders (VAEs). In contrast with the broad use of LLMs across M\&S stages, the review demonstrated that these \textit{other GenAI techniques have been used in fewer studies, covering fewer stages}. These techniques have primarily been used to generate data as input to models (e.g., population synthesis), particularly with GANs~\shortcite{montevechi2021input} and VAEs~\shortcite{sane2025comprehensive}, while less common applications have used GANs for model validation~\shortcite{montevechi2022using} or to design experiments~\shortcite{feldkamp2022method}. 

So far, there is a paucity of M\&S studies using GenAI to create images. This absence can be surprising for at least three reasons. First, there is a clear awareness of GenAI for image (or video) synthesis, in part due to high-profile societal and legal concerns such as deepfakes~\shortcite{ma2025social} and their risks for media manipulation~\cite{farid2025mitigating}, or potential copyright infringements~\cite{coulter2024aiming}. Second, many text-to-image generation systems are closely related to GenAI techniques that have been used in the M\&S space for several years: text-to-image generation pipelines typically rely on VAEs, while post-processing and image transformation tasks use GANs. Third, text-to-image generation no longer requires training from scratch or access to specialized hardware. It has reached a level of technical maturity and accessibility, as evidenced by modern systems that are available through stable software libraries or user-friendly portals and provide reproducible workflows. Given its broad visibility, methodological continuity with VAEs and GANs already used in M\&S, and increasing technical maturity, why has text-to-image generation remained largely absent from M\&S studies?

Two factors may help explain this limited uptake. First, \textit{the applicability of text-to-image generation to any stage of the M\&S process is not sufficiently clear}. Conceptual modeling can rely heavily on textual descriptions, while implementation emphasizes executable code, both of which align with LLMs much more than image generation. Simulation outputs also typically take the form of structured data rather than images. Second, even where potential applications exist, \textit{practical guidance on how to use text-to-image generation within the M\&S context is lacking}. Although the underlying technology is mature and widely accessible, the ecosystem comprises multiple model families, interfaces, and workflows, so there is no one-size-fits-all solution that modelers can readily use. As a result, identifying appropriate tools and integrating them meaningfully into M\&S practice can present a nontrivial barrier. The main contribution of this tutorial is to address these two barriers by showing modelers the potential applications of text-to-image generation at different stages of the M\&S process, and supporting these applications through an examination of the tools and options from a M\&S perspective.

This tutorial is structured into two main parts. Section~\ref{sec:why} motivates \textit{why} text-to-image generation can have a place in the M\&S process by examining four use cases: \textit{explaining} a model by generating illustrations of its mechanisms, supporting \textit{empathetic decision-making} by seeing what happens to simulated entities (e.g., the `virtual lives' of people), enabling \textit{hybrid simulation} by using images as an interface between different types of models, and generating \textit{training material}. Building on these use cases, Section~\ref{sec:how} examines \textit{how} to perform text-to-image generation, exposing the choices faced by modelers (e.g., which GenAI model, what embeddings) and relating options (e.g., conditional control) to their specific needs. For instance, a photorealistic GenAI model is appropriate to immerse viewers in a simulation, but may not be the right fit when illustrating conceptual models. Rather than scripts or step-by-step instructions tied to a specific text-to-image model, this tutorial emphasizes transferable workflows and decision principles so that modelers gain the skills to select and adapt appropriate GenAI tools as the technology changes.

\section{New and Emerging Applications of Text-to-Image Generation in M\&S}
\label{sec:why}

\subsection{Tailored Communication: Supporting Model Explanation with Text-to-Image Generation}
\label{sec:explanations}

Modelers are rarely their own customers: they typically work in interdisciplinary teams together with model commissioners, subject-matter experts (SMEs), and other stakeholders. Communication therefore plays an important role throughout the model building and assessment process, as well as in supporting the future uptake and acceptability of a model. Among several authoritative references, the AQuA book emphasizes that \textit{high-quality analysis must be clearly communicated} and accepted by its commissioners in order to be fit for purpose, and that engagement with stakeholders across the analytical life cycle contributes to trust, relevance, and effective decision support~\cite{aqua_book}. While implementations, code, or standardized documentation (e.g., ODD+D for Agent-Based Models) are appropriate forms of communications among modelers, public-facing communication or interdisciplinary teamwork often requires additional forms of support. In a manifesto on models that (should) serve society by~\shortciteN{saltelli2020five}, several modelers pointed out the necessity of tailored, accesible communication:
\begin{quote}
    ``Simplified, plain-language versions of the model can be crucial. When a model is no longer a black box, those using it must react to assess individual parameters and the relationships between them. This makes it possible to communicate how different framings and assumptions map into different inferences, rather than just a single, simplified interpretation from an overly complex model. Or to put it in jargon: qualitative descriptions of multiple reasonable sets of assumptions can be as important in improving insight in decision makers as the delivery of quantitative results.''
\end{quote}

Although we often think of text, and early GenAI studies on communicating models were exclusively text based~\shortcite{shrestha2022automatically}, a series of WSC tutorials on conceptual modeling by Robinson has repeatedly exemplified that visuals can play a role. Most recently, in quoting Ronald N. Giere, he pointed out that a conceptual model is a representation and that its format ``can be words, equations, diagrams, graphs, photographs, computer-generated images''~\cite{robinson2025tutorial}. For example, modelers may engage with SMEs through workshops by \textit{sketching conceptual models} together to clarify key constructs and mechanisms, and modelers may later demonstrate the adequacy of a model by \textit{showing how its outputs compare to expectations}, such as visual comparisons of time series or spatial patterns of incidence.

Text-to-image generation can be as simple as using services such as Google's Gemini or OpenAI's ChatGPT to describe a situation (e.g., states and transitions of a model) and get a diagram, thus serving as a convenient aid. However, our focus is on \textit{automation} and using GenAI as part of the M\&S process. In this context, we stress that communication is about addressing a specific audience with the aim of conveying a particular message, so being effective requires \textit{providing information in forms that align with the needs, expectations, and levels of expertise of the audience}. The ability to adapt by generating visuals on-the-fly is where GenAI tools can shine. Consider a complex problem such as suicide prevention: it would be overwhelming for a modeler to manually compile six different reports addressed to psychiatrists, epidemiologists, health economists, policymakers, families, and individuals with lived experiences of suicide. Besides, even within a single stakeholder category such as policymakers, there are national roles and informational needs between national, state, or county-level authorities. However, LLMs can automatically transform conceptual models and/or simulations into reports based on the needs and preferences of an audience~\cite{giabbanelli2025towards}, and text-to-image generation tools can automatically create illustrations for these reports. Alternatively, text-to-image generation can populate a slide deck, and voice synthesis software can turn it into a presentation. Such materials could contribute to broadening access to simulations~\cite{giabbanelli2024broadening}: instead of models being black boxes or the information being conveyed in a generic report, end-users can access reports based on their specific concerns.

Concretely, the process needs to specify \textit{what} should be illustrated, \textit{where} visuals should be placed, and \textit{how} they should be produced. In order of difficulty, the location of visuals is the easiest aspect to address: we can assume that visuals appear after the text they help illustrate, similarly to floating content in \LaTeX. Determining how to produce visuals raises two related questions: \textit{what prompt} should be used, since text-to-image generation starts from a prompt, and \textit{which GenAI tool} should render it. A simple dichotomy can help structure this choice: are we visualizing abstractions, or are we aiming for photorealism? Visualizing abstractions such as processes and structures typically calls for an illustrative style, which can be obtained through tools such as OpenAI's Dall-E. By contrast, visualizing specific situations, outcomes, or lived experiences can benefit from photorealistic representations that convey detailed and context-rich environments. Many tools support photorealism, as shown in Section~\ref{sec:choosing}.

The prompt is a more complex matter. We start with the content to be illustrated (e.g., a paragraph in a report describing part of a model), then modifications should account for three considerations. First, the prompt may need to be adapted to the expected \textit{format} of the GenAI tool: some tools perform better with lists of keywords, while others respond more effectively to fully formed sentences. Since both formats are often accepted as input, a modeler may not realize when they use a sub-optimal format. Second, there is the issue of \textit{guardrails} to avoid the generation of problematic outputs~\shortcite{ma2026safety}. Strong guardrails in some tools (e.g., OpenAI's Dall-E) can hinder work on triggering topics; for instance, a tool may refuse to illustrate a report on suicide when prompts include keywords related to self-harm, requiring the modeler to rephrase the prompt. Conversely, tools with weaker safety mechanisms may generate inappropriate content, so modelers may need to add guardrails directly within the prompt. Third, prompts must reflect the \textit{communication standards of the field}. For example, guidelines for media portrayals of individuals affected by obesity~\cite{obesity} recommend avoiding visuals that ``depict individuals affected by obesity engaging in stereotypical behaviors (e.g., eating junk food, engaging in sedentary behavior)'' unless accompanied by counter-portrayals such as healthy eating or physical activity. Similarly, when illustrating reports on suicide prevention, guidelines advise to ``avoid any imagery that portrays the person in pain or sadness or shows means of suicide''~\cite{suicideMedia}. Because GenAI tools cannot be assumed to internalize such domain-specific standards, modelers may need to encode these expectations explicitly in their prompts. From an automation perspective, the most challenging issue remains identifying \textit{what} to illustrate. For instance, if a workflow limits visuals to one per paragraph, selecting which idea would benefit most from visualization remains an open-ended problem. As an initial approach, algorithms may identify the most central sentence of a paragraph, although this does not guarantee that the resulting visual adds value rather than redundancy.

\begin{figure}[h]
    \centering
    \includegraphics[width=1\textwidth]{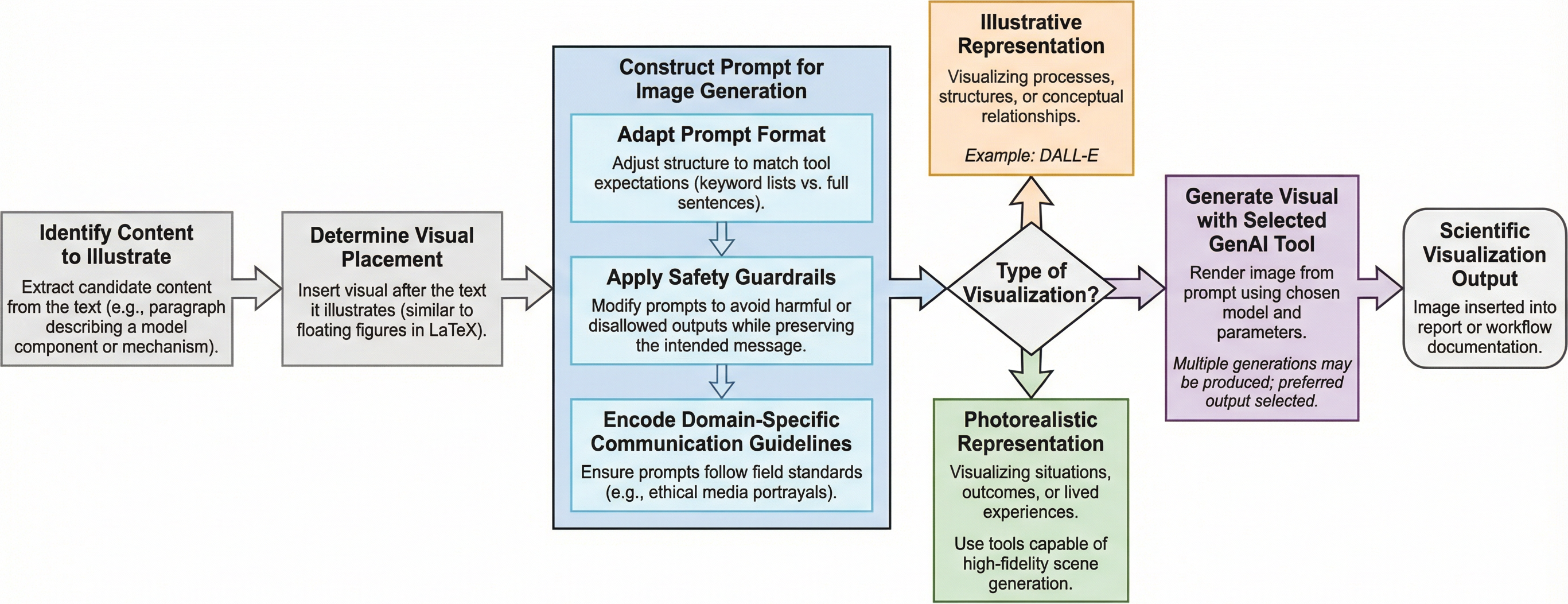}
    \caption{This visualization of the options for text-to-image generation was automatically produced from the paragraph in section \protect\ref{sec:explanations} by using OpenAI's GPT to set the design (text-to-text) then Google's Nano Pro for rendering (text-to-image). The text is cleanly rendered (no typos), boxes and arrows are straight.}
    \label{fig:abstractProcess}
\end{figure}

To demonstrate text-to-image generation, we used the following prompt for GPT: ``Your task is to write a detailed description of a visual, such that it can be rendered independently by another LLM. This is a scientific visualization of a workflow, suitable for publications in a peer-reviewed article. Be clear about the location of boxes, use of arrows, and so on. Colors (if any) should be used to reinforce information, rather than just to give different colors to boxes. The process that you need to visualize is described as follows.'' We copied and pasted the two paragraphs above. Then, we used Google's Nano with the prompt ``produce a 16:9 wide image with a purely white background as a scientific illustration for the visual described as follows'', giving it the output from GPT. The model being stochastic, we generated images three times and selected our preferred result, shown in Figure~\ref{fig:abstractProcess} as an illustration of the process. This approach combines GPT's ability to plan and describe together with the quality of Nano Pro for rendering processes.

\subsection{Bringing Simulation Outputs to Life Through AI-Generated Visuals}

Communicating simulation models and their results to decision-makers and stakeholders is a longstanding M\&S challenge, often addressed by producing statistics, charts, or maps. However, aggregate outputs (e.g., number of evacuees reaching safety or average travel time for an evacuation simulation) can obscure the experiences of individuals represented in a simulation, so decision-makers may not fully appreciate the human consequences of policies and interventions. Several approaches have sought to reduce the divide between policymakers and the populations represented in simulations by translating model outputs into more accessible forms. Recent work has shown that simulation results can be conveyed through narratives that summarize the overall behavior of a system~\cite{flandre2026distilling} or recount the experiences of individual agents~\shortcite{giabbanelli2025promoting}, thereby helping readers interpret complex dynamics and consider the consequences of interventions from the perspective of those affected. While these textual approaches improve accessibility and can promote empathetic reasoning, they still rely on readers to mentally construct the situations being described. Visual representations offer a complementary pathway: by depicting the environments and actors involved in a simulation, \textit{images can make modeled scenarios more immediate and concrete, reducing the cognitive effort required to imagine the consequences of a policy or event}. Furthermore, psychological research suggests that imagery plays a central role in how people perceive risks. For example, experimental evidence shows that enhancing mental imagery associated with environmental hazards increases emotional engagement and leads individuals to perceive climate-related risks as more severe~\shortcite{karlsson2023causal}. In practice, if we run workshops with stakeholders on managing forest fires or flooding, seeing flooded streets or damaged neighborhoods can be much more persuasive than visualizing a spread on a cellular automaton or discussing aggregate measures. 

Text-to-image generators enable the rapid creation of photorealistic scenes that could help convey the outcomes of simulations. However, simulation outputs cannot typically be used as prompts for image generators directly. Most simulations produce numerical outputs (e.g., counts of number of cells in a certain state, average value of agents over time) that lack the contextual information needed for a text-to-image model to construct a coherent scene. For example, a cellular automaton modeling flood propagation may report the number of flooded cells or the maximum flooded area~\shortcite{zellner2025enhancing}, but such values do not specify what the environment looks like, what type of buildings are present, or what conditions characterize the event. Consequently, \textit{simulation outputs must first be translated into textual descriptions that can serve as prompts for image generation}. In many cases, this translation can remain relatively simple. \textit{Simulation outputs can populate structured templates} (see Listing 1) describing the situation to be visualized. For instance, if a simulation indicates that several streets are flooded or that a neighborhood experiences severe flooding, these quantities can be inserted into a prompt describing a flooded urban environment. Although this step introduces an intermediate representation, the transformation from simulation outputs to prompts remains minimal because it simply converts quantitative results into short textual statements.

\begin{lstlisting}[language={},caption={Text-to-image template to generating a visual from outputs of a hypothetical Agent-Based Model (ABM) for evacuation. Simulation-derived fields are shown as placeholders. Results are in Figure \protect\ref{fig:evacuation}.},label={lst:abm_prompt}]
Photorealistic documentary-style image illustrating an evacuation during a flood.
Agent attributes from simulation: An (*@\placeholder{AGE\_GROUP}@*) (*@\placeholder{GENDER}@*) evacuating from their home. The agent belongs to household type (*@\placeholder{HOUSEHOLD\_TYPE}@*) and is currently in evacuation state (*@\placeholder{EVACUATION\_STAGE}@*). The agent is moving toward the nearest safe zone after receiving the evacuation warning.
Simulation context: Location: (*@\placeholder{NEIGHBORHOOD\_TYPE}@*) neighborhood in a (*@\placeholder{CITY\_SIZE}@*) town. Flood conditions: floodwater depth approximately (*@\placeholder{WATER\_DEPTH\_CM}@*) cm on the street. Weather conditions: heavy rain following (*@\placeholder{RAIN\_DURATION\_HOURS}@*) hours of rainfall.
Environment: Residential streets with houses and parked vehicles partially surrounded by floodwater. Water is muddy and flowing slowly
along the street. The sky is dark with storm clouds and the atmosphere reflects an ongoing emergency situation.
Agent action: The person is walking through the flooded street while evacuating the area, moving carefully through the water toward higher ground.
Visual style: Realistic photojournalism, eye-level perspective, natural lighting, dramatic storm atmosphere.
\end{lstlisting}

\begin{figure}[h]
    \centering
    \includegraphics[width=1\textwidth]{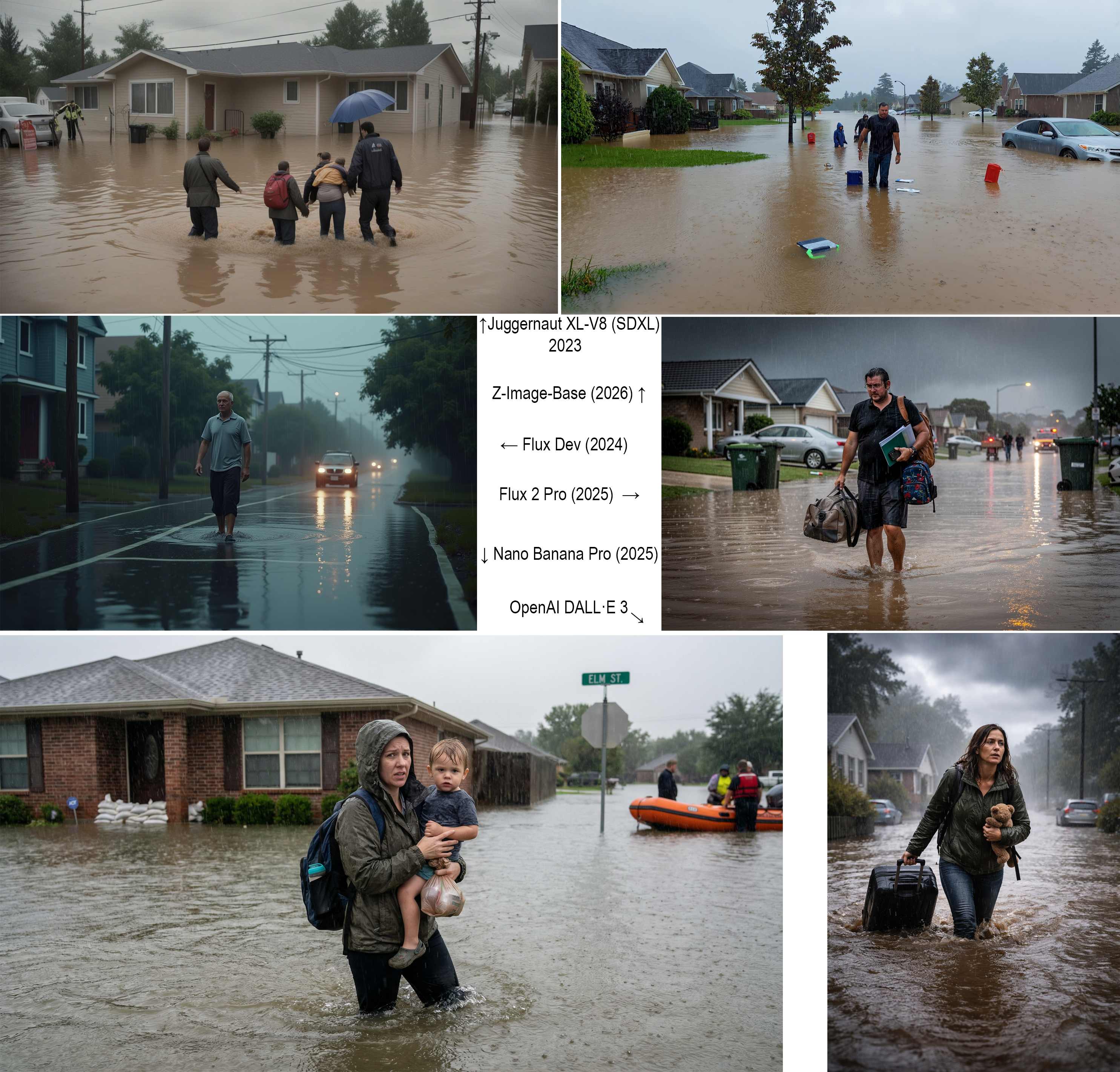}
    \caption{We populated Listing 1 with agent attributes (adult, single parent, stage `moving to shelter') and generated using six models. Since ABMs work with demographic variables and states rather than visuals (e.g., agent hair color or clothing), visual aspects are not fully specified so repeated generations with the same prompt may produce different visuals (e.g., the agent may change hair color). Key technical differences between models include `early generators' (e.g., SDXL) with many anatomical glitches (hands, faces) and current tools (e.g., Nano) that are anatomically correct and correctly embed text in the image. \textit{This high resolution composite image can be zoomed in. Original files are on our repository.}
    }
    \label{fig:evacuation}
\end{figure}

However, basic prompts may not always provide sufficient guidance, depending on the text-to-image generators and on the application. Without adequate contextual information, some generators may produce generic environments or aesthetically pleasing scenes that fail to reflect the intended situation. For instance, we may produce a beautiful image of a pristine town, on a sunny day, with turquoise waters covering the streets as if it was a resort. This would fail to meet expectations as a visual for the weather and degraded conditions associated with a flood. In these cases, we envision three options. First, we can work on the prompt to make it as complete as possible so that little is left for the generator to guess: the type of environment, the weather conditions, the perspective, and the cues that convey the severity of the event. One way to construct such detailed prompts is to introduce an intermediate LLM, preferably with reasoning capabilities and potentially fine-tuned on the application context. \textit{This LLM could take in results from the simulation and craft a detailed text description of the scene}, so that the generated image faithfully represents the situation implied by the simulation results. 

Second, we can \textit{ground the generation process with images}. Some generators accept images as inputs alongside text prompts, allowing the user to constrain the visual output more tightly. In this case, images can serve different purposes. For example, a photograph of a specific location can be provided so that the generator modifies that environment according to the simulation results (e.g., showing a familiar street under flood conditions). Alternatively, example images can illustrate the visual characteristics of the phenomenon of interest, such as the appearance of streets or houses after flooding. Image-guided approaches have been explored to communicate environmental risks by superimposing simulated flood levels onto photographs of real locations, producing visuals that resemble the places affected while conveying projected hazards~\cite{siegel2021superimposing}. Such techniques help ensure that generated visuals remain anchored in recognizable environments and realistic conditions rather than drifting toward generic or aesthetically optimized scenes. By combining textual prompts derived from simulations with reference images, modelers can exert greater control over the generated visuals and better align them with the phenomena being simulated.

A third option is to \textit{correct the image} after generation, i.e. post-processing method through image-to-image techniques. Generative Adversarial Networks (GANs) are particularly well-suited for this task. For example, \shortciteN{schmidt2019visualizing} created CycleGAN to learn the mapping between a picture of a normal street and a flooded environment. Because the geometry of the original image is preserved, the resulting visuals remain anchored in recognizable locations while depicting simulated impacts. We stress that a GAN is not intended to directly render a simulation: it performs visual post-processing by translating an image into a target domain.

Accepting elements of fiction to become immersed (in a simulation) is the notion of `suspension of belief'. The viewer knows that these are generated images from simulations, not events happening to real people or places. The viewers may accept this artificiality as long as it maintains internal plausibility. \textit{When something inconsistent appears (e.g., the same character suddenly looks different), the suspension of disbelief is broken, which immediately disrupts immersion}~\cite{green2000role,lombard1997heart}. This is a concern when using text-to-image generators with simulations. For example, in an ABM, consider that a user clicks on an agent to visualize their situation, leading the system to generate a photorealistic depiction of that individual (e.g., a blond woman evacuating her neighborhood). If the user clicks later on the same agent but the generator produces a different appearance (e.g., a red-haired woman leaving the area), then there is an issue. The same problem arises for environments: if a simulation explores several interventions for the flooding of a specific village, the visuals should depict the same location under different conditions. \textit{Text-to-image generators do not guarantee continuity}. They generate new samples from a distribution of plausible images~\shortcite{rombach2022latent}. Locking the random seed can reproduce the same image when the prompt is unchanged\footnote{When using a third-party LLM, different users may get different results based on unexpected factors such as their session history or location in the world. However, replicability is not a problem for text-to-image generation: if another user is given the same workflow, prompt, and seed, then they will get the same image (unless their workflow uses an LLM itself).}, but small prompt modifications (e.g., the same village is now described under flood conditions) may alter the appearance of the scene. Ensuring consistency across generated visuals requires more advanced techniques~\shortcite{avrahami2024chosen,wang2025characterfactory}, particularly when repeatedly visualizing the same agents or locations under evolving conditions.

\subsection{Generating Visual Representations for Human and Machine Learners}

Researchers and educators emphasize the importance of real-life projects or, when these are impractical, realistic case studies~\shortcite{van2010panel}. However, educators struggle to access such teaching cases~\cite{de2019simulation}. Text-to-image generation could help address this shortage of instructional material by producing visual case studies on demand, with the desired properties. For example, an instructor teaching conceptual modeling may elect to start with sparse acyclic graphs, then move onto more complex dense feedback-rich models. This is not only a matter of generating \textit{structures}, which could already be done with text-based LLMs as they can output a graph (e.g., as a list of edges). In conceptual modeling, the representation itself matters. The tutorial by \citeN{robinson2025tutorial} stresses the value of conceptual models to communicate with clients and other participants in a simulation study. A `good' conceptual model does not only have a sound structure: it is also visually organized based on standards of the field and supports interpretation. For example, it would be strange if a System Dynamic model had stocks drawn in trapezoids (instead of square boxes), and it would be bad practice if concepts were haphazardly located so that many edges are unnecessarily crossing or labels overlap. Text-to-image generation could thus help to \textit{ensure that educators have a wealth of (controllable) case studies, rendered in a professional target style} (e.g., conventions of the M\&S environment such as Vensim, style of expert-produced models). In that sense, generated visuals could support homework, examinations, and adaptive practice by giving students many more examples than would normally be feasible to prepare manually.

A second opportunity lies in training computational systems. In introductory modeling and simulation courses, images are often used to explain phenomena such as diffusion, congestion, or emergent spatial structure. A classic example is to show how a process such as forest fire spread depends on modeling choices~\cite{karakonstantis2026review}, for instance whether the environment is represented with a square grid or a hexagonal lattice. If such visuals can be generated in a controlled manner, then they can be varied systematically across structural assumptions, parameter values, and visual encodings. This would \textit{create large collections of aligned examples from which a multimodal LLM could learn simulation concepts}. In a recent study, we `taught' modeling and simulation to GPT-4o by providing it with the complete slide decks of the course as visuals~\cite{flandre2024can}. Access to this course material improved performance, and GPT-4o achieved passing grades across modeling themes and levels of mastery, often matching or exceeding a median student.

\subsection{Interfacing between models}
Text-to-image generation can serve as an interface between heterogeneous models in multi-scale simulation workflows. For example, ABMs can simulate how individuals move to an airport and how diseases may propagate based on exposure~\shortcite{han2023disease}, while physics-based models such as computational fluid dynamics can simulate aerosol transport and ventilation effects~\shortcite{vuorinen2020modelling}. In such cases, attributes generated by the ABM could be used to produce images or geometric representations consistent with an agent’s characteristics (e.g., face shape and mask type), which are then used as inputs for the fluid dynamics model (e.g., to compute the risk to inhale and exhale particles), thus updating the agent's infection risks~\cite{giabbanelli2025emerging}.

A related opportunity arises in biomedical multi-scale modeling, where population-level simulations interact with models operating at the organ or tissue scale. ABMs are widely used to study infectious disease dynamics in structured environments such as hospitals, schools, or public buildings. At finer scales, however, disease progression and diagnosis often rely on imaging-based representations such as radiographs or MRI slices. In principle, generative models could produce medical images conditioned on agent attributes (e.g., age, sex, BMI, or comorbidities), enabling downstream models or classifiers to estimate disease progression or treatment outcomes for simulated individuals. Although current general-purpose text-to-image models are not trained on sufficiently representative medical datasets for such applications, specialized generative models and privacy-preserving training strategies may eventually make this interface feasible.

\section{How to Use Text-to-Image Generators from an M\&S Perspective}
\label{sec:how}
\subsection{Choosing a text-to-image generation model: which characteristics impact M\&S tasks}
\label{sec:choosing}

Selecting an appropriate text-to-image generator for M\&S tasks depends on several characteristics that differ substantially across models. Table~\ref{tab:modelList} exemplifies these characteristics as of June 2026, but technology changes quickly and other factors (e.g., license, or costs for cloud-based solutions) can play a role. Unlike artistic or casual uses of image generation, M\&S may have reproducibility requirements and the need to generate large numbers of images from structured simulation outputs. Practitioners should therefore consider factors such as the \textit{style of prompts} expected by a model (see subsection~\ref{sec:howToPrompt}), computational requirements (e.g., how much memory is needed), \textit{generation speed}, and deployment constraints. We consider running a model locally as the more desirable option because it supports reproducibility, privacy, and integration within simulation pipelines. However, the feasibility of local deployment varies considerably across models. Loading a model may require anywhere from a few gigabytes of GPU memory to several tens of gigabytes\footnote{The GPU memory is not a hard requirement in the sense that a computer with less than 12 Gb of RAM cannot run a generator requiring 12 Gb. Rather, when RAM is insufficient, several strategies allow the image generation to still proceed. For example, tiled or chunked decoding techniques divide the image into smaller regions that are processed sequentially to reduce peak memory usage. Some inference frameworks also support offloading parts of the model to the operating system’s swap space, effectively extending available memory beyond the GPU. These solutions should be considered as `back-up' plans because they typically increase generation time and may introduce additional overhead from memory transfers.}. In addition, inference (i.e., the steps that make the image) vary, as some of them may require dozens of steps, whereas newer architectures aim to produce comparable outputs with fewer iterations.

\begin{table*}[t]
\centering
\caption{Comparison of selected text-to-image generators relevant for modeling and simulation workflows. Nano Banana Pro is the only cloud-based model; all other models have open weights and can be downloaded to run locally on consumer-grade computers, which also allows to control the random seed for consistency. Early models (Stable Diffusion, SDXL) had limited prompt adherence, notable anatomical issues, and struggled to render text in an image; these shortcomings are usually addressed with modern models. The table shows a sample of models, as many more are available (e.g., Qwen-Image, Imagen 4, ERNIE).}
\small
\begin{tabular}{p{2cm}p{2cm}p{2cm}p{2cm}p{2cm}p{2cm}p{2cm}}
\hline
\textbf{Characteristic} & \textbf{Flux 1 Dev (2024)} & \textbf{Flux 2 Klein 9B (2025)} & \textbf{Stable Diffusion 1.5 (2022)} & \textbf{SDXL -- Juggernaut v9 (2023)} & \textbf{Nano Banana Pro (2025)} & \textbf{z-Image Base (2025)} \\
\hline
Parameter scale & $\sim$12B & $\sim$9B & $\sim$1B & $\sim$3--6B & undisclosed & $\sim$6B \\ \hline

Typical VRAM requirement & 16--24 GB & 12--16 GB & 6--8 GB & 12--16 GB & cloud only & $\sim$12 GB \\ \hline

Inference speed & moderate & very fast (few steps) & moderate & moderate & fast (cloud optimized) & fast \\ \hline

Prompt style expected & natural language descriptions & natural language descriptions & comma-separated keyword tags & natural language sentences & instruction-style prompts & long caption-style descriptions \\ \hline

Prompt length tolerance & high & high & limited & medium & high & very high \\ \hline

Photorealism quality & very high & very high & moderate & high & very high & very high \\ \hline

Human anatomy accuracy & strong & strong & weaker & strong & strong & strong \\ \hline

Fine-tuning support & limited & limited & extensive (LoRA ecosystem) & extensive & none & limited \\ \hline

Typical resolution & 1024+ & 1024+ & 512 native & 1024 native & high resolution & 1024+ \\ \hline

Strengths for M\&S & detailed scene rendering & fast large-scale generation & lightweight local deployment & good semantic reasoning & multimodal reasoning & detailed scene descriptions \\ \hline

Weaknesses for M\&S & smaller tooling ecosystem & fewer tools & weaker prompt reasoning & heavier compute & not locally deployable & newer ecosystem \\

\hline
\end{tabular}
\label{tab:modelList}
\end{table*}

\subsection{How to Prompt a Generator}
\label{sec:howToPrompt}

\textit{The behavior of a text-to-image model reflects the characteristics of the data on which it was trained}: what images it has been exposed to, and what was the style of the associated prompts. These models often rely on large-scale sets of (image–text) pairs from the web. For example, the LAION-5B dataset contains billions of image–text pairs collected from Common Crawl and filtered to retain pairs where the textual description roughly corresponds to the visual content~\shortcite{schuhmann2022laion}. Such training regime explains why diffusion models respond well to descriptive language grounded in \textit{observable visual features} (e.g., an adult man holds his head between his hands and looks away while tears roll from his eyes to his cheeks), while abstract or underspecified prompts (e.g., unhappy image of the man not feeling well) may lead to unpredictable outputs. Prompting styles vary because models are trained on different types of image–text datasets and captioning conventions\footnote{Several tools can help practitioners learn prompt writing conventions by reversing the generation process. Instead of crafting prompts from scratch, modelers may provide example images for what they seek (e.g., flood evacuation, forest fire, crowd simulations) and use automatic taggers or captioners to infer the tags or prompts that would likely reproduce similar visuals. For example, the \textit{Danbooru Tags Transformer} combined with the \textit{WD Tagger} (available as a public HuggingFace Space) analyzes an uploaded image and produces structured tag lists or prompts suitable for text-to-image models such as Stable Diffusion. These tools are trained to recognize visual attributes (e.g., objects, poses, environments, or stylistic cues) and return standardized tags that correspond to the vocabulary commonly used in diffusion prompting~\shortcite{wang2023diffusiondb}. Practically, a modeler who wants to generate images for an ABM of hospital operations could upload reference photos of crowded emergency departments to learn tags related to spatial layout, lighting, and human activities. By observing how existing visuals are decomposed into prompt elements, they can progressively build prompt templates, reducing trial-and-error.}. Stable Diffusion was trained on captions derived from web metadata so they are short descriptive phrases or keyword-like annotations, resulting in the common practice of comma-separated lists of concepts for Stable Diffusion 1.5 and SDXL (e.g., ``photorealistic portrait, adult man sitting, emotional distress,  looking away, sorrowful expression, dramatic lighting, cinematic mood, documentary photography, detailed face, natural skin texture, shallow depth of field''). More recent models have moved toward richer textual supervision, so they work with fully formed sentences. Understanding how models react to prompts is an active research topic. Despite guidelines from developers, analyses of large collections of real prompts reveal that effective prompting often involves iterative experimentation and the use of stylistic modifiers or keywords that emerged from community practice~\shortcite{wang2023diffusiondb}.

Pragmatically, we view text-to-image generation as a complement to existing M\&S workflows, supporting specific tasks such as those outlined in Section 2. So instead of expecting modelers to develop yet another area of expertise in prompt engineering for image generation, it may be more practical to rely on processes and tools that improve or partially automate prompt formulation. Such processes would let modelers focus on the simulation tasks. \textit{Prompt rewriting approaches} attempt to translate abstract user requests into more concrete descriptions of objects, attributes, and scenes, which target models can interpret given their training data~\shortcite{fan2024prompt}. The literature on prompt optimization also makes a useful distinction between `hard prompts' (plain text: a sequence of human-readable words) and `soft prompts' (embedding vectors that cannot be read by users). As explained in the next subsection, a text-to-image model does not directly operate on a text prompt: it converts it into embeddings using an encoder. So, technically, the output of a simulation does not have to turn into text in order to use an image generator: it could be directly turned into embeddings, which can be learned through optimization~\shortcite{wen2023hard}. Note that we still recommend text prompts for many M\&S tasks because they are directly interpretable and portable across image generation models, which change very frequently. Regardless of the quality of the prompt, modelers should be mindful that prompt compliance and generation is ultimately limited by the model: complex scenes, spatial relationships, or multi-object descriptions are challenging~\shortcite{sridhar2024adapting}. 

To appreciate the differences between prompting style, consider the following case study. \citeN{anagnostou2013distributed} developed an emergency medical services (EMS), which integrates an agent-based model (ABS) of ambulance services  with discrete-event models (DES) of hospital accident and emergency. Agents dispatch vehicles, transport patients, and select hospitals. The internal processes of hospital departments account for triage and treatment queues. The simulation produces operational outputs such as ambulance response times, numbers of patients waiting in triage or treatment queues, hospital bed occupancy, staff availability, ambulance arrivals, which serve as key performance indicators for EMS management. In the paper, outcomes are visualized through histograms of ambulance response times and distributions of patient journey duration. For the purposes of this tutorial, consider that we want to generate images corresponding to different simulated outputs, e.g. a crowded emergency department when bed occupancy and patient queues are high. Figure~\ref{fig:evacuation} shows characteristics of generators: SDXL has anatomical glitches and EMS are less busy if the prompt uses fully formed sentences, Nano can cope either way and generates correct text. While the cloud-based Nano has the best images, zImage and Flux run locally.

\begin{figure}[h]
    \centering
    \includegraphics[width=1\textwidth]{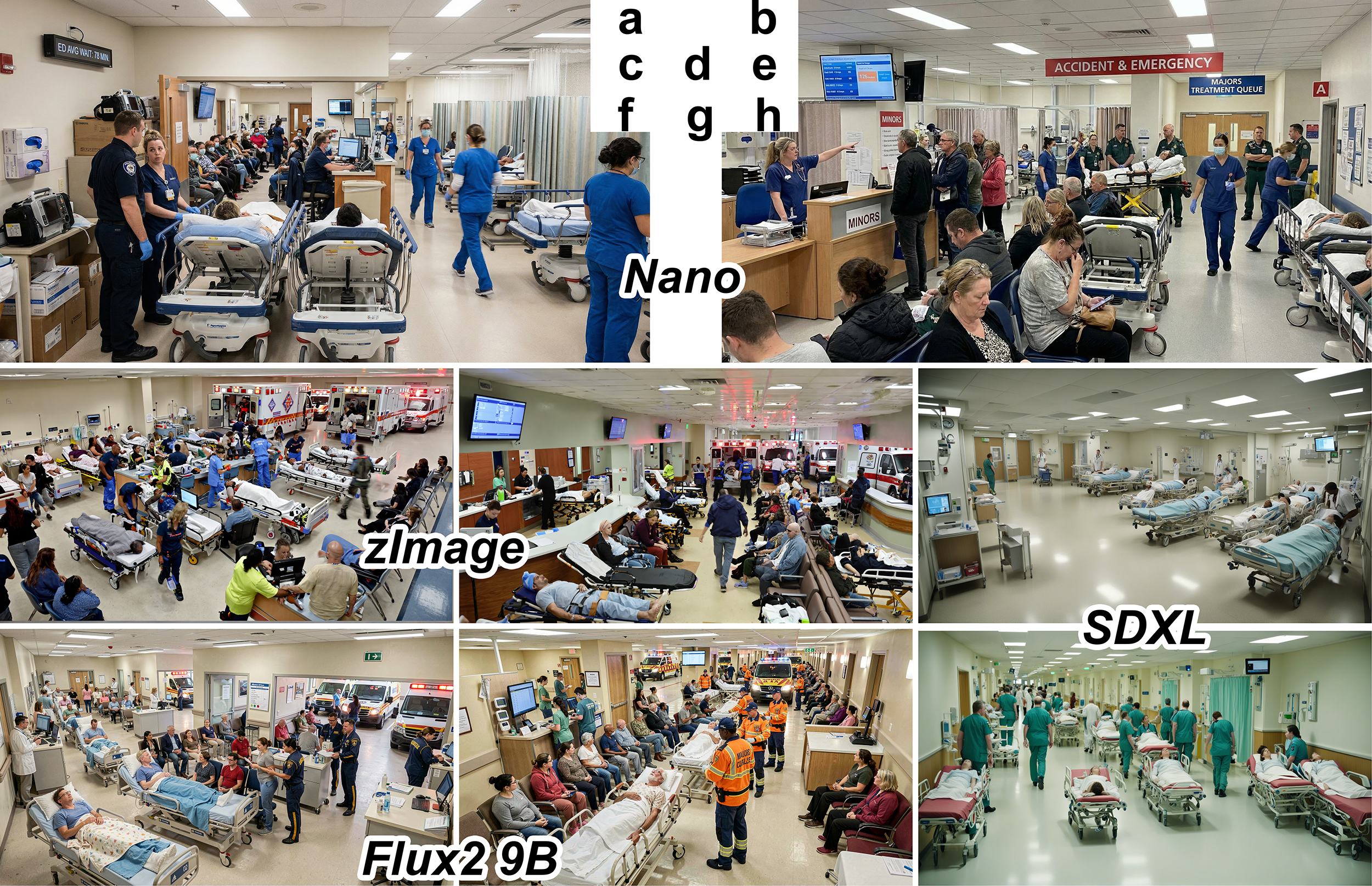}
    \caption{For four generators, we show a pair of outputs where the first (left or top) uses a long narrative prompt and the second (right or bottom) uses a list of comma-separated tags. Generators such as Nano and Flux2 (a, b, f, g) can cope with either format, while older generators like SDXL (e, h) do better with tags and generators trained on detailed captions may produce richer images when given longer narratives (c, d). \textit{This high resolution composite image can be zoomed in. Original files are on our repository.}}
    \label{fig:hospital}
\end{figure}

\subsection{Running Models Locally: Core Elements of Text-to-Image Models}

Software such as ComfyUI, Forge, and A1111 can execute \textit{workflows} locally to produce images on consumer-grade computers. These workflows are portable between software as they are coded as JSON files. In addition, software allow for the complete workflow to be embedded in the images: for example, a drag-and-drop of an image with embedded workflow into ComfyUI will automatically load the workflow and notify the user if any custom components need to be downloaded and installed to replicate the workflow. These features enable provenance and replicability, which are important requirements for M\&S. Our permanent repository at \url{https://doi.org/10.5281/zenodo.19545751} includes all high-resolution individual images generated in this tutorial alongside with embedded workflows (for locally generated images).

\begin{figure}[h]
    \centering
    \includegraphics[width=1\textwidth]{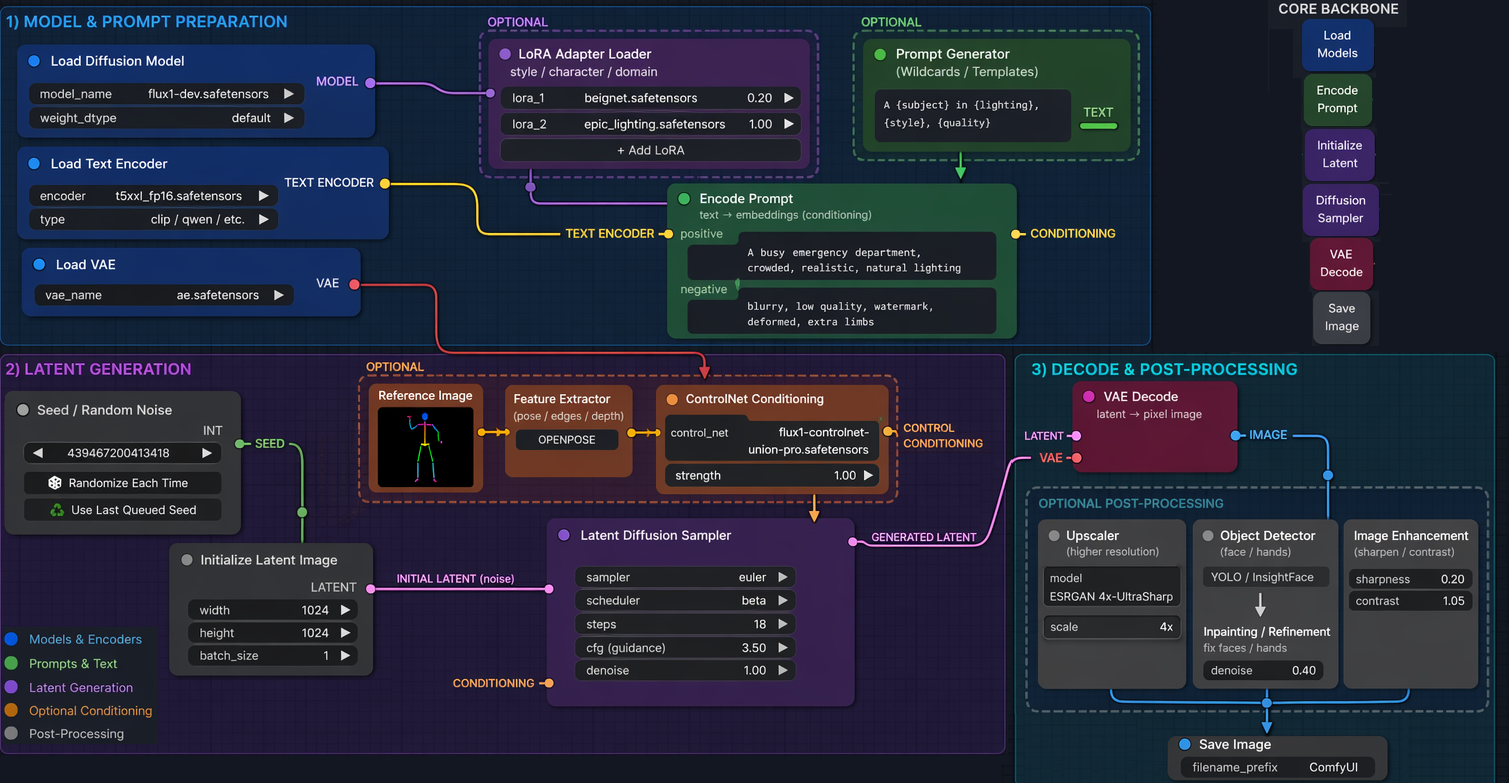}
    \caption{This high-resolution figure in the style of a ComfyUI workflow summarizes the core of the process and the options available to modelers. The process starts with step \#1 in the top-left.}
    \label{fig:workflow}
\end{figure}

While such workflows may appear visually complex, they have the same core sequence of operations: loading models, encoding the prompt, generating an image in latent space through an iterative denoising process, and finally decoding the result into a visible image (Figure~\ref{fig:workflow}). We briefly explain these steps to support M\&S in reusing workflows by understanding the purpose of each main component. First, we need to load three elements\footnote{After these models have been loaded, we can optionally load lightweight adapters known as LoRAs (Low-Rank Adaptation modules) which can modify style or characters and ensure consistency. Since we cannot re-train a model such as Flux or z-Image but we have access to its weights, the main way to `customize' them is to train LoRAs. The AI Toolkit can be used to create such LoRAs from pairs of images and captions, and large databases of LoRAs are available on a variety of websites such as tensor.art, civitai.com, civarchive.com, tungsten.run, or mage.space. However, these websites are primarily for artists and/or recreational users, so they are low on scientific assets (e.g., hospital layouts) and primarily contain characters.}: the model (e.g., Flux, Stable Diffusion), the variational autoencoder (VAE), and one or more text encoders. Earlier models such as Stable Diffusion XL typically packages several components together in a single `checkpoint', while more recent workflows load the components independently. 

Next, the user's prompt is processed by the text encoder to produce embeddings that guide the generative process. Historically, many Stable Diffusion workflows relied on dual prompts: a \textit{positive prompt} describing what should appear in the image and a \textit{negative prompt} specifying features to avoid (e.g., artifacts or undesirable styles). More recent models often rely on a single prompt, since the undesirable features (e.g., blurry, bad anatomy) are less common. Prompts can also be automated or parameterized, e.g. using wildcards/placeholders that are replaced with randomly selected elements from predefined lists. These placeholders are \textit{not} intended to be replaced by simulation outputs. Rather, they introduce controlled visual variety while keeping the simulated situation unchanged. For instance, in three hospital congestion scenario where the simulation indicates that patients are waiting in an overcrowded emergency department, the prompt may always describe the same operational conditions (e.g., patients waiting on stretchers, limited staff), while placeholders vary secondary elements such as camera perspectives (eye-level view, view from the triage desk, corridor perspective).

The generation does not directly happen on a full-resolution image like a 1024x1024 grid of pixels. Rather, it operates on a compressed representation of the image, called a \textit{latent representation}, produced by the VAE. This latent space has far fewer dimensions than the final image, e.g. a 1024×1024 image may be represented internally as a much smaller tensor describing abstract visual features such as shapes, textures, and spatial structure. Generation begins with a \textit{latent image} containing random noise. The diffusion model then performs an \textit{iterative denoising process}\footnote{Several parameters control this denoising process, including the number of denoising steps, the guidance scale (CFG) that determines how strongly the prompt influences generation, and the denoise strength controlling how much noise is removed during each iteration. Note that research continues to improve these generation processes, for example with CFG-Zero*~\shortcite{fan2025cfg} and CFG-Ctrl~\shortcite{wang2026cfg}, so modelers can rely on the fact that there is a guidance mechanism but its specific implementation may change over time. \textit{These parameters are very model dependent}: SDXL models may have a CFG from 5 to 9 (typically 6 or 7) while Flux models are in the 2-4 range (typically 3.5). The choice of sampling algorithm (e.g., Euler, Heun, DPM variants) and scheduler also affects the trade-off between speed and image quality. Because the process is stochastic, workflows usually include a random seed parameter that determines the initial noise pattern. Fixing the seed allows the same image to be reproduced under identical conditions, whereas randomizing the seed produces new variations.}: at each step, the model predicts how the noisy latent representation should be slightly adjusted so that it becomes more consistent with the prompt. In other words, the model repeatedly removes a small amount of noise while adding structure guided by the prompt embeddings. After several steps, the latent representation gradually evolves from random noise into a coherent latent image. Finally, the VAE decoder converts this latent representation in pixel space. Optional post-processing steps may then refine the output. These can include sharpening filters, super-resolution upscalers (e.g., the open source Real-ESRGAN), or targeted correction stages. For example, the anatomical issues encountered with earlier models were often fixed by using detection models such as Ultralytics' Yolo series~\cite{hidayatullah2026yolo26,schuerkamp2023enabling} to detect hands and faces, then applying \textit{in-painting} (i.e., regenerating a selected region of the image while keeping the rest unchanged) to fix the content. 

The prompts and parameters of the generation process are not the only ways to control the generation. Optionally, a ControlNet allows external signals (e.g., poses extracted with OpenPose, edge maps, or depth estimates) to constrain the spatial structure of the generated image. In such cases, an auxiliary image is analyzed to extract a structural representation, which is then provided as an additional input during diffusion. For instance, in a hospital congestion scenario, the simulation might provide the layout of beds, stretchers, and waiting patients in an emergency department. A ControlNet conditioned on a layout map or segmentation mask could ensure that the generated image preserves the simulated spatial organization of the ward while the text prompt controls the visual style (e.g., lighting, medical equipment, or staff activity).


\section*{ACKNOWLEDGMENTS}

The author thanks Dr. Megan Witherow (Old Dominion University) for several discussions that have benefited the direction of this tutorial. The author also thanks current and former students (Noé Flandre, Tyler Gandee), whose work on text-to-image generation and multimodal LLMs provided valuable examples.

\footnotesize

\bibliographystyle{wsc}

\bibliography{demobib}

\section*{AUTHOR BIOGRAPHIES}

\noindent {\bf \MakeUppercase{Philippe J. Giabbanelli}} is a Research Professor at Old Dominion University in Norfolk, Virginia. He received a B.S. in Computer Science from the University of Nice (France) and holds M.S. and Ph.D. degrees from Simon Fraser University (Canada). He has held positions at several universities (e.g., the University of Cambridge) prior to being tenured at Miami University and then promoted to the Virginia Modeling, Analysis, and Simulation Center (VMASC). He has authored over 180 articles (mostly on Modeling \& Simulation and AI applied to human behavior), of which 15 have appeared at the Winter Simulation Conference, where he co-leads the professional development track. His e-mail address is \email{pgiabban@odu.edu}.\\

\end{document}